\pdfoutput=1
\RequirePackage{ifpdf}

\documentclass{JINST}

\usepackage{cite}

\title{A Simulation of the Optical Attenuation of TPB Coated Light-guide
Detectors}

\author{B.J.P. Jones$^a$\thanks{Corresponding Author}\\
\llap{$^a$}Massachusetts Institute of Technology,\\
  77 Massachusetts Avenue, Cambridge, MA 02139, United States of America\\
  E-mail: \email{bjpjones@mit.edu}}

\abstract{This note is provided as a supplementary section to accompany the
paper \cite{ChristinaProceedings} which has been included in these proceedings.
It describes some simple simulations which were performed in order
to understand the attenuation behaviors of acrylic light-guides operated
in air and argon, which were characterized in \cite{Baptista:2012bf}. Whilst these simulations are only at the level of
sophistication of a toy model, they illustrate interesting non-exponential
light attenuation effects and the differences between operating light-guide based detectors in argon
and air environments. We investigate the effects of surface absorption,
surface roughness and wavelength dependence, and use a model tuned
on the light-guide attenuation curve measured in air to make a prediction of the light-guide attenuation curve in
argon.  This curve is compared with data from a liquid argon test stand, and an improvement over
a simple exponential model is observed.}


\begin{document}

\section{Introduction and Motivation}

A series of tests of coated acrylic light guide detectors for future
liquid argon TPC detectors have been made in both air and argon \cite{Baptista:2012bf}.
Both cast and extruded acrylic bars were tested in air, and these have different surface
finishes and transverse dimensions, and hence different attenuation behaviors. 
In the first round of tests, only cast acrylic 
bars were tested in argon.  During analysis of the attenuation data taken in argon and air, 
two unexpected features emerged. First,
the attenuation length measurement obtained from the two sets of tests
were not in good agreement. Second, the attenuation behavior
in argon was not observed to obey a simple exponential decay.

Whilst initially these features were thought to be caused by
instrumental effects in the test stands, subsequent analysis shows
that these effects should be expected, to some extent. This note describes
a toy simulation which illustrates the origin of both the argon vs
air attenuation length discrepancy and of nonexponential attenuation. 
The model has a single free parameter, corresponding
to the coefficient of surface absorption, which we tune using data
taken in air and use to make a prediction of the attenuation curve
in argon. Despite being a significant
oversimplification of the real test geometry, which involves several
different surfaces with different absorption coefficients, the prediction
made for argon is a signficant improvement over a simple exponential
attenuation curve when compared with data. Future studies of coated and uncoated light-guide
surface properties could be used to tune a more advanced model with
stronger predictive power.

\section{Description of the Ray Tracing Model}

To model the expected light attenuation behavior in acrylic light
guides operated in argon and air we have performed a series of simple
ray tracing simulations. In these simulations, isotropic light rays are generated,
and each ray is assigned a weight determined by the attenuation processes under consideration.  This weight represents the probability that a photon following
this trajectory will arrive at the detection end of the light guide, where there is a coupling to a photomultiplier tube or silicon photomultiplier.  The relative
detected light yield is given by the sum of the weights
of all the isotropic rays emitted from a wavelength shifting surface.  The bar is represented
by a region of solid acrylic with refractive index 1.49, in an environment
of either air with refractive index 1.0, or argon with refractive
index 1.23. The critical angle for total internal reflection is given
by 

\begin{equation}
sin\theta_{c}=\frac{n_{environment}}{n_{acrylic}}
\end{equation}

In principle, light which impinges upon a surface at an angle larger
than this will have a reflection coefficient of 1, and light at lower
angles will experience a partial reflection with a calculable coefficient
$R(\theta)$. In our model, light rays are emitted by the TPB coating
isotropically, and for those rays emitted into the bar we calculate
the angle of reflection against each set of surfaces. The number of
reflections is then given by

\begin{equation}
N_{i}=\frac{1}{tan\theta_{i}}\frac{L}{w_{i}}
\end{equation}

Where L is the length of the bar, $\theta_{i}$ is the angle of the
ray to the surface normal and $w_{i}$ is the width of the bar in
this direction. The index i labels the two sets of surfaces.

We can imagine three main sources of light attenuation: 
\begin{enumerate}
\item Loss of light rays from partial reflection below $\theta_{c}$
\item Absorption of light in the acrylic bulk
\item Absorption of light rays at each bounce due to an imperfect optical
surface
\end{enumerate}
We account for (1) by applying an attenuation factor $\alpha_{1}$
to each ray which is below the critical total internal reflection
angle, where 

\begin{equation}
\alpha_{1}=\prod_{i}R(\theta_{i})^{N_{i}}
\end{equation}
The factor $R(\theta)$ is the reflection coefficient expected for
a light ray impinging upon an ideal surface separating two materials
of different refractive indices, averaged over polarizations. 

Additional corrections to this factor may be expected to occur from
surface roughness, which causes small changes to the angle of incidence
and reflection with each bounce, and therefore allows light rays to
deviate outside the region of total internal reflection and escape.
We modeled the effect of random surface deviations of between 0.01
and 0.1 radian per bounce, and found a very minimal effect upon the
attenuation curves. This is because such small deviations only cause
significant losses for light which is very near the critical angle,
which in practice is a negligible fraction of all detectable light
rays. Surface roughness on the scale of the wavelength of light requires
a more in depth treatment, and has not been investigated.

Previous measurements of bulk attenuation in many commercially available acrylics at 440~nm give attenuation lengths of
several meters \cite{MiniCleanAcrylic}.  This suggests that the 40~cm attenuation observed
in our light guides is likely to be primarily due to the surface losses
of factor (3).  Therefore, we assume that the effects of bulk absorption are negligible compared to surface effects.  This assumption could be easily relaxed if more information on the bulk or surface properties of our acrylic become available.  For the purposes of this model, we set

\begin{equation}
\alpha_{2}=1
\end{equation}

Factor (3) is incorporated into our model by reducing the weight
of each light ray by a constant coefficient of surface absorption,
$R_{S}$ for each reflection. Previous studies of acrylic light guides
\cite{HuffmanTPB} suggest that $R_{S}$ is in the few percent range. However,
we expect the numerical value of $R_{S}$ to depend heavily on the
manufacturing process and surface finish of the particular acrylic
bars used, and so we treat it as a free parameter in the model.

\begin{equation}
\alpha_{3}=(1-R_{S})^{\sum_{i}N_{i}}
\end{equation}

The total attenuated weight of each ray is then given by $\Pi_{j}\alpha_{j}$,
and the total light output of the bar is given by the sum of the weights
of many randomly generated isotropic light rays.

\subsection{Effective Absorption Length from Surface Losses \label{sub:EffectiveAbsorption}}

\begin{figure}[tb]
\begin{centering}
\includegraphics[width=0.8\columnwidth]{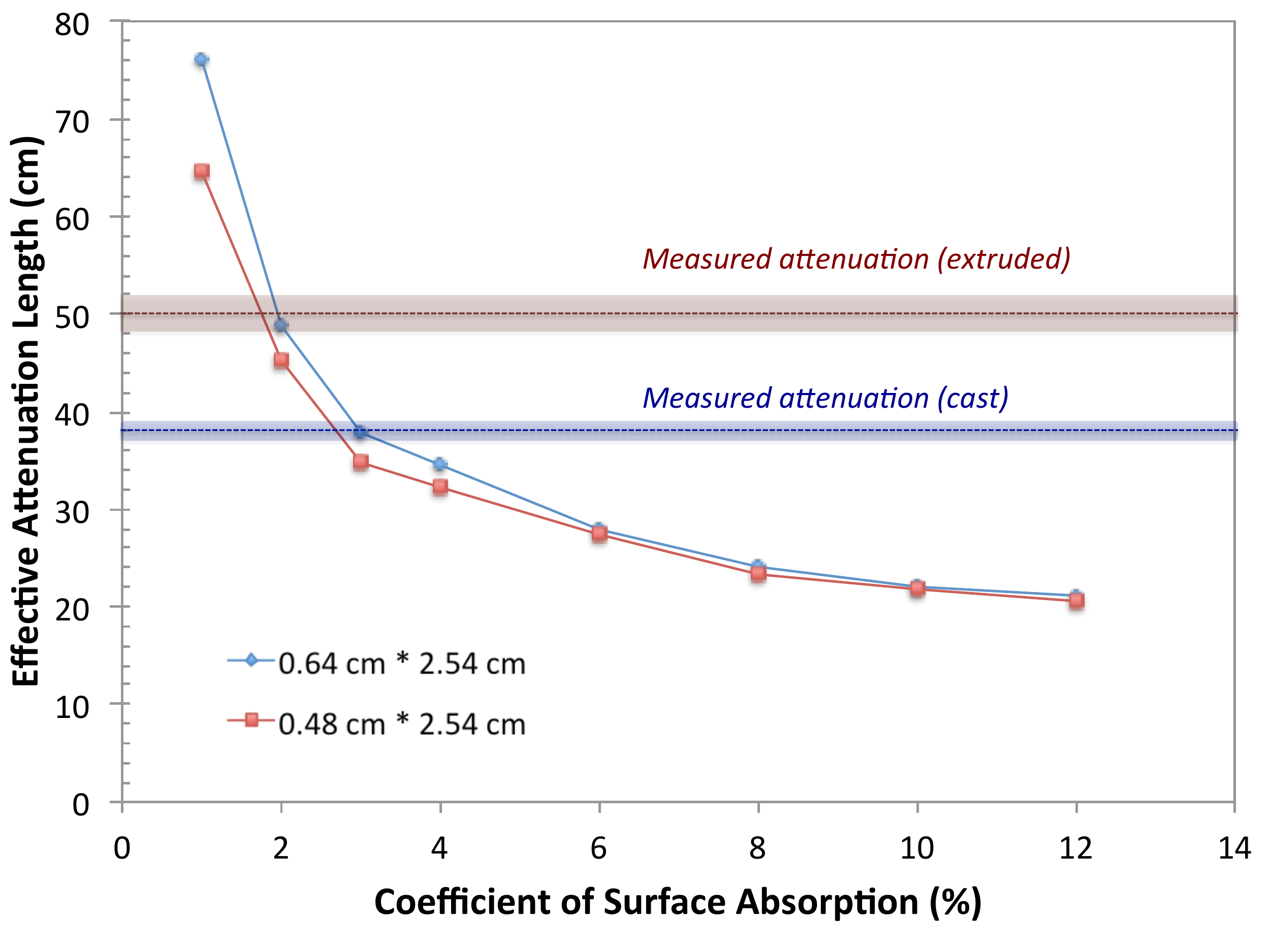}
\par\end{centering}

\caption{Calculated attenuation due to surface absorption for different light
guides. Measurements from \cite{Baptista:2012bf} are overlaid.  A full description is given in the text. 
\label{fig:Calculated-attenuation-due}}
\end{figure}

Assuming an infinite bulk absorption length, we can calculate the
the effective attenuation length in the 20-50~cm region for different
surface absorption coefficients. The actual attenuation length generated
by surface absorptions has a dependence on the transverse bar dimensions,
since a narrower bar leads to more bounces per unit length than a
wider bar. 

We have simulated the effective 20-50~cm attenuation length for both
0.48x2.54~cm bars and 0.64x2.54~cm bars in air for several values of the
coefficient of surface absorption. The calculated attenuation lengths are shown
in Figure \ref{fig:Calculated-attenuation-due}. If all attenuation
in our light guides were from surface absorption, we expect a per-bounce
absorption coefficient of around 2\% for the extruded acrylic bars (which have the smaller cross section),
and around 3\% for cast acrylic bars (which have the larger cross section).

\subsection{Non-exponential Form of Attenuation Curves}

The form of the attenuation curve produced by our model is not expected
to be exponential. Since both the number of surface reflections per
unit length and the subcritical ray reflection coefficient depend
on the angle of the ray being traced, rays at different angles are
lost at different rates. Therefore the angular composition of the
light beam is changed at larger distances, and the attenuation behavior
has a nontrivial distance dependence. The attenuation behavior also
depends on the refractive index of the operating environment, since
the critical angle and subcritical ray reflection coefficients both
have a refractive index dependence. Hence we expect to see different
attenuation behavior in argon and air. 

\begin{figure}[tb]
\begin{centering}
\includegraphics[width=0.8\columnwidth]{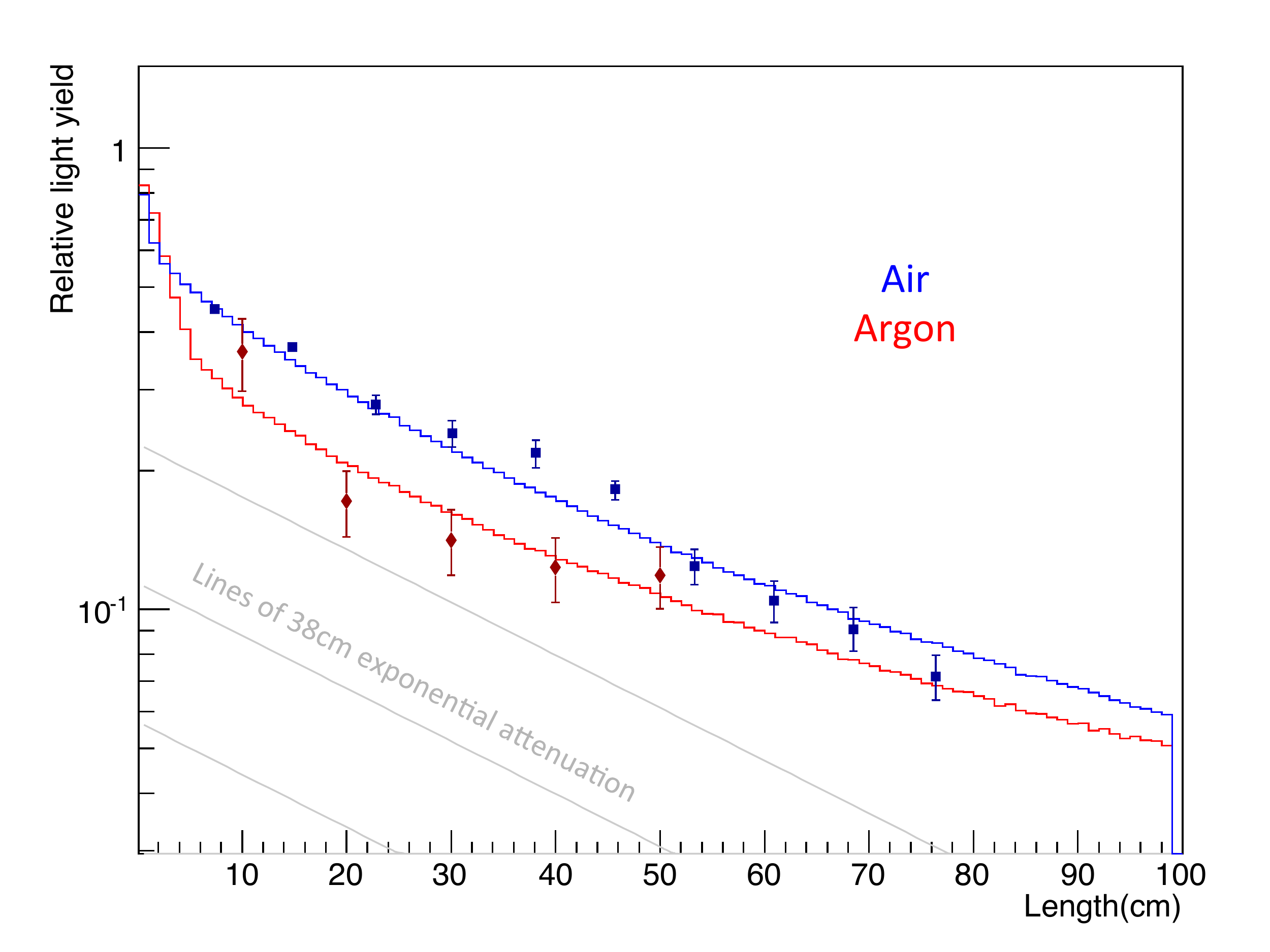}
\par\end{centering}
\caption{Calculated attenuation curves assuming identical bars operated in
argon and air, with 3\% surface absorption per reflection and a long
bulk attenuation length. Data from \cite{Baptista:2012bf} are overlaid. \label{fig:Calculated-attenuation-curves}}

\end{figure}

Assuming that the cast acrylic rods used in both the argon and air
studies have identical indices of refraction and surface absorption
coefficients, we can calculate the shapes of the attenuation curves
given an infinite bulk attenuation length. These curves are shown,
with measurements from  \cite{Baptista:2012bf} overlaid, on figure \ref{fig:Calculated-attenuation-curves}.
On the same plot are marked lines of 38~cm exponential attenuation
to highlight the nontrivial attenuation behavior. The only tunable
parameter in this simulation is the per-reflection absorption coefficient,
which is set to 3\%, as extracted from the effective attenuation length
of the cast acrylic light guides in air as described in section \ref{sub:EffectiveAbsorption}. 

\subsection{Incorporating Wavelength Dependent Effects}

\begin{figure}[tb]
\begin{centering}
\includegraphics[width=0.7\columnwidth]{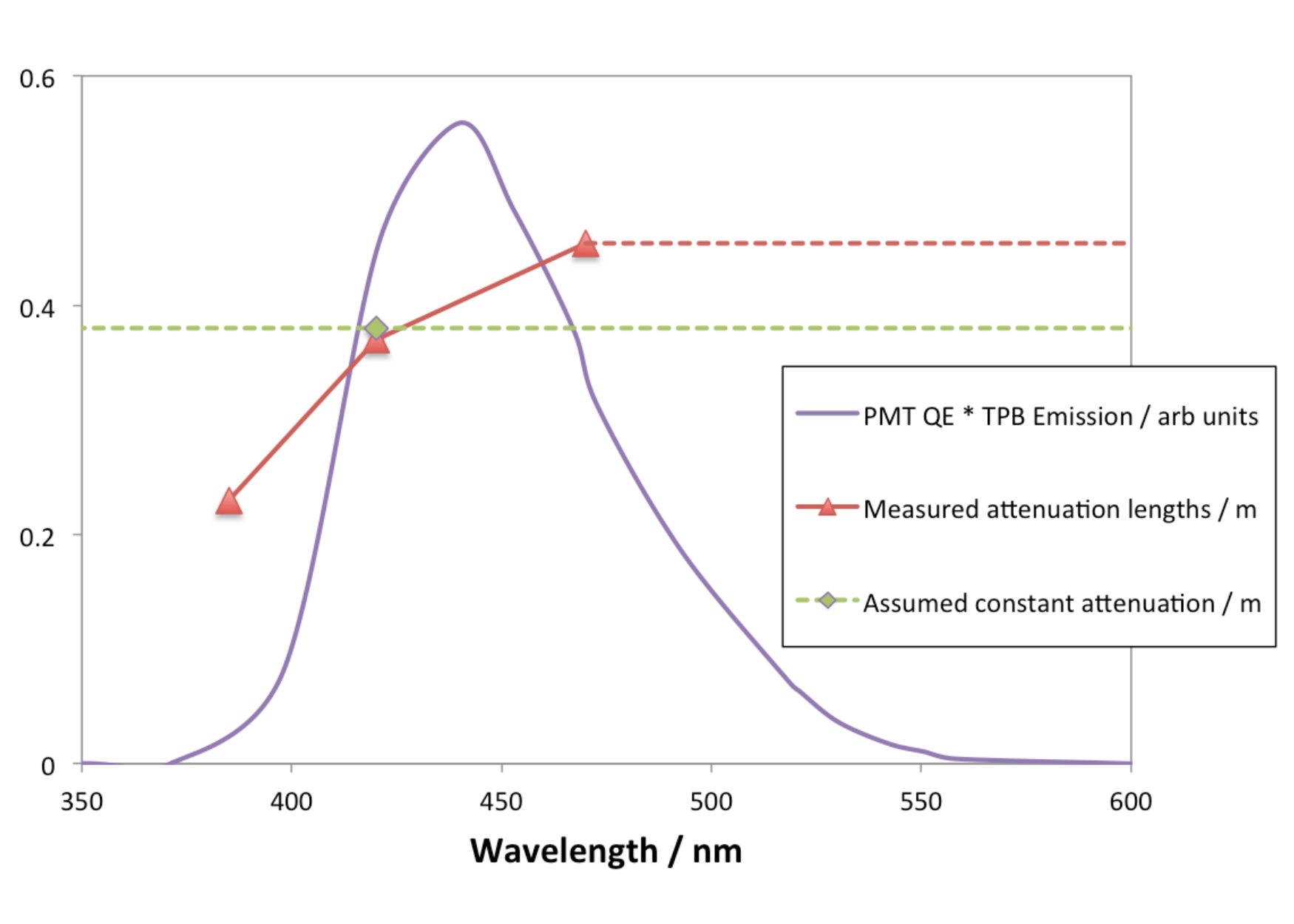}
\par\end{centering}

\caption{Measured wavelength dependence over relevant spectral range (red),
compared to previously assumed constant attenuation (green). Extrapolated ranges with no data are shown as dashed lines.
\label{fig:Measured-wavelength-dependent}}
\end{figure}

Measurements of attenuation of monochromatic light at the University
of Indiana indicate that TPB coated light-guide attenuation has a wavelength dependence
over the TPB emission spectral range. The measured wavelength dependent
attenuation length is shown in Figure \ref{fig:Measured-wavelength-dependent}
overlaid on a curve showing TPB emission spectrum multiplied by PMT
quantum efficiency, which gives the relevant spectrum of detectable
rays. There is a 10-20\% variation over the range of interest, leading
to contributions both above and below the previously assumed constant
attenuation length of 38~cm. We expect this variation to produce an
additional nonexponetnial contribution to the attenuation behavior.

\begin{figure}[tb]
\begin{centering}
\includegraphics[width=0.8\columnwidth]{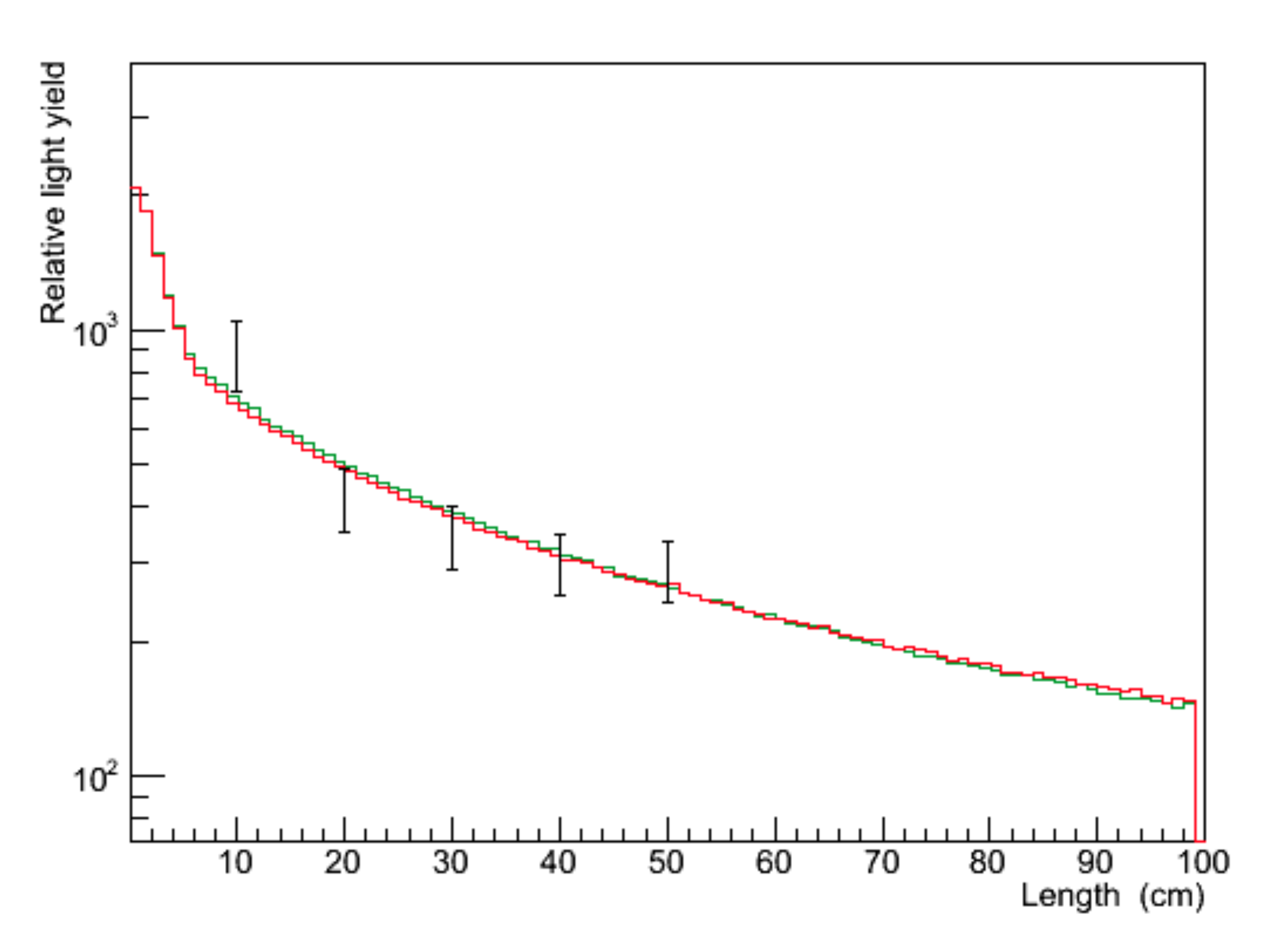}
\par\end{centering}

\caption{Nonexponential attenuation generated by wavelength dependent (red)
and wavelength independent (green) ray tracing models 
\label{fig:Nonexponential-attenuation-gener}}

\end{figure}

Incorporating these effects into the ray tracing simulation described
above, we can produce an attenuation curve with a wavelength dependent
surface absorption. This curve is shown overlaid on the previously
generated curve, which assumed a constant absorption coefficient,
in Figure \ref{fig:Nonexponential-attenuation-gener}. We see that
the overall effect averaged over all detectable rays is minimal, mainly
because there are both positive and negative contributions at different
points in the spectral range. We have also investigated the effect
of this wavelength dependence on the simple bulk attenuation model,
and found its impact to be similarly minimal.

\acknowledgments
This work was supported by the National Science Foundation (PHY-1205175).

\bibliographystyle{JHEP}
\bibliography{LightguideSimulations.bib}{}

\newpage

\end{document}